\documentclass{emulateapj}
\usepackage{color} 
\usepackage[colorlinks,urlcolor=blue,citecolor=blue,linkcolor=blue]{hyper ref}
\usepackage{epsfig}
\usepackage{subfigure}
\usepackage{graphicx}
\usepackage{commath}
\usepackage{amsmath}
\usepackage{natbib}
\usepackage{mathtools}

\begin{document}

\title{Turbulent-diffusion Mediated CO Depletion in Weakly Turbulent Protoplanetary disks}
\author{Rui Xu\altaffilmark{1}, Xue-Ning Bai\altaffilmark{2}, Karin \"{O}berg\altaffilmark{3}}

\altaffiltext{1}{Department of Astrophysical Sciences, Princeton University, Princetion, NJ 08544;
ruix@princeton.edu}
\altaffiltext{2}{Institute for Theory and Computation, Harvard-Smithsonian Center for Astrophysics, 
60 Garden St., MS-51, Cambridge, MA, 02138; xbai@cfa.harvard.edu}
\altaffiltext{3}{Harvard-Smithsonian Center for Astrophysics, 
60 Garden St., MS-16, Cambridge, MA, 02138; koberg@cfa.harvard.edu}

\begin{abstract}
Volatiles, especially CO, are important gas tracers of protoplanetary disks (PPDs).
Freeze-out and sublimation processes determine their division between gas and solid phases, which affects both which disk regions can be traced by which volatiles, and the formation and composition of planets.
Recently, multiple lines of evidence suggest that CO is substantially depleted from the gas in the outer regions of PPDs, i.e. more depleted than would be expected from a simple balance between freeze-out and sublimation.
In this paper, we show that the gas dynamics in the outer PPDs facilitates volatile depletion through turbulent diffusion.
Using a simple 1D model that incorporates dust settling, turbulent diffusion of dust and volatiles, as well as volatile freeze-out/sublimation processes, we find that as long as turbulence in the cold midplane is sufficiently weak to allow a majority of the small grains to settle, CO in the warm surface layer can diffuse into the midplane region and deplete by freeze-out.
The level of depletion sensitively depends on the level of disk turbulence. Based on recent disk simulations that suggest a layered turbulence profile with very weak midplane turbulence and strong turbulence at disk surface,
CO and other volatiles can be efficiently depleted by up to an order of magnitude over Myr timescales.
\end{abstract}

\section{Introduction}\label{sec:intro}

Protoplanetary disks (PPDs) consist of gas and dust. Both components
play a major role in planet formation through dynamical processes in the
gaseous disk, as well as physical and chemical coupling between gas and
dust components. The dust can be probed via the disk spectral energy distribution
and resolved dust continuum emission up to millimeter/centimeter grain sizes
\citep{Andrews15}. Despite uncertainties in dust opacity, dust
mass can be derived from sub-millimeter continuum flux (e.g.,
\citealp{WilliamsCieza11}). There is no corresponding direct constraint on the gas
because molecular hydrogen hardly
radiates, the gas mass is instead usually estimated by assuming a canonical
gas-to-dust mass ratio of 100 from the interstellar medium (e.g.,\citealp{Bohlin_etal78}),
leading to large uncertainties.

Recently, a number of works have attempted to measure the gas
content of PPDs using CO and its isotopologues (e.g.,
\citealp{Bruderer2012,WilliamsBest14,Kama2016a,Kama2016,Eisner_etal16,Ansdell_etal16}).
As a volatile species, CO freezes out onto dust grains in the cold midplane regions
of the outer PPDs, while it remains in the gas phase in the warmer disk surface layer
(e.g., \citealp{HenningSemenov13}).
These studies, which incorporate CO freeze-out and different levels of disk chemistry,
found that if one assumes a standard
gas to dust ratio and a canonical $\rm CO/H_2$ ratio of $\sim10^{-4}$ (e.g.,
\citealp{Frerking_etal82,Ripple2013}), CO is frequently underabundant by 
a factor of $\gtrsim10$ in the warm disk surface layer.
This result holds also if isotopologue-selective photodissociation is taken into
account (\citealp{Miotello14,Schwarz16}).
Therefore, either CO is intrinsically depleted, or the gas-to-dust mass
ratio is significantly lower than the standard value.

\begin{figure*}[!ht]
\centering
\includegraphics[width =0.8\textwidth]{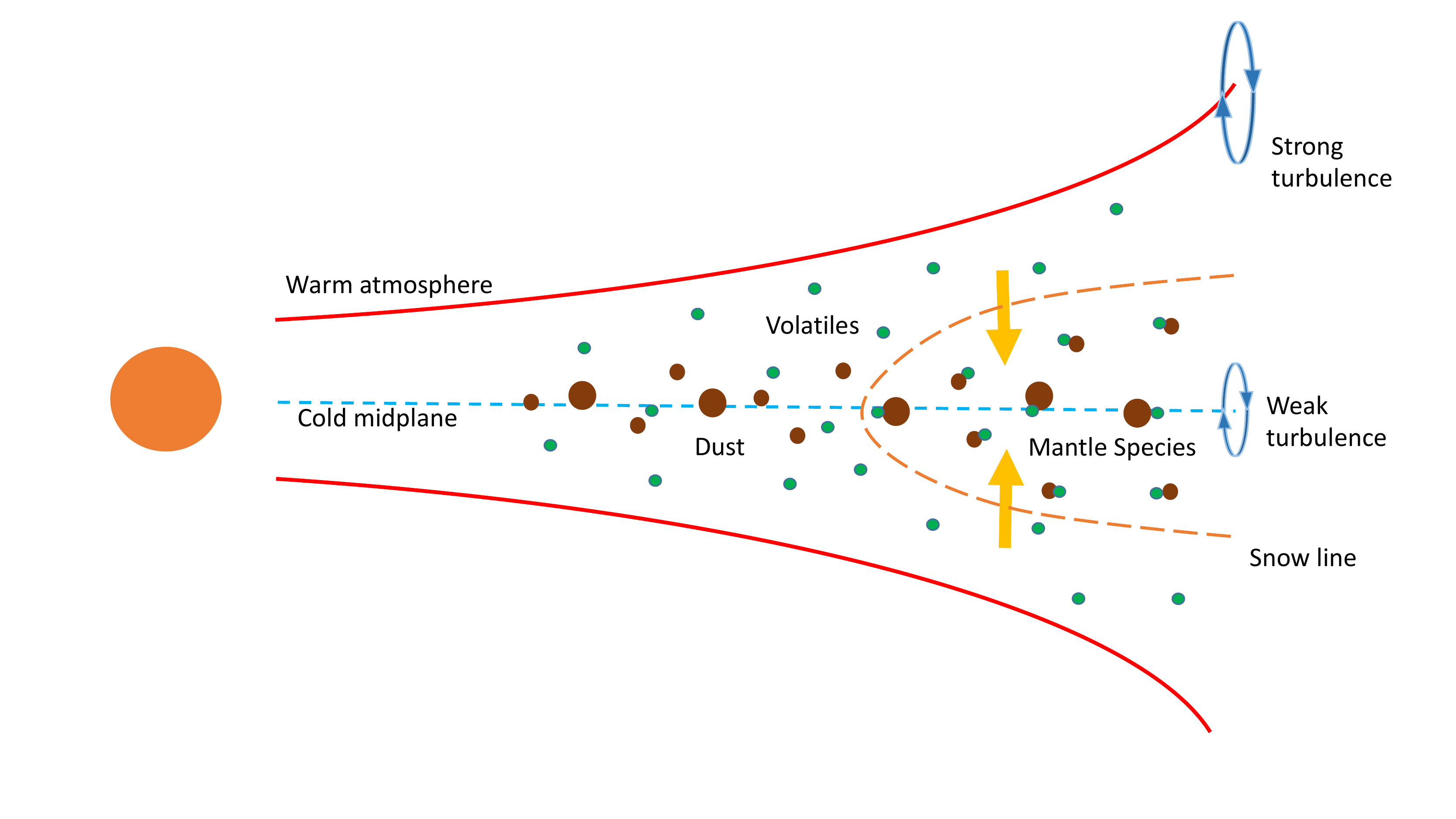}
\caption{Schematic picture on the turbulent diffusion mediated volatiles (CO)
depletion in PPDs.
Freeze-out of CO on dust grain surface at the low-temperature midplane allows
surface CO to turbulently diffuse down to the midplane, which further freeze-out
onto the grains. Dust grains settle to the midplane without mixing back to disk
surface due to weak midplane turbulence, leading to systematic CO depletion.}
\label{fig:PPDs}
\end{figure*}

Theoretically, both scenarios are plausible. The gas-to-dust ratio can
be reduced via disk wind, where mass loss from disk surface primarily
remove gas instead of dust \citep{Gorti_etal15,Bai16}. In the mean time,
through chemical processes, a significant fraction of carbon can be
converted to complex organic molecules over the disk lifetime
\citep{Bergin2014,Yu2016,Bergin_etal16}. The presence of CO depletion is
supported at least in the case of TW Hya \citep{Favre_etal13,Du_etal15},
where a constraint on the
disk gas mass is available from HD observations \citep{Bergin_etal13}.

Volatile depletion has also been inferred in the case of water, whose
freeze-out temperature is much higher.
Based on {\it Spitzer} mid-infrared observations of H$_2$O lines
\citep{Salyk08,CarrNajita08}, \citet{Meijerink09} showed that water vapor
abundance at the disk surface is sharply truncated beyond $\sim1$AU,
inconsistent with pure chemical models (e.g. \citet{Glassgold09}). They
hypothesized that beyond $\sim1$AU, warm water vapor at the disk surface
diffuses vertically towards the midplane by turbulence and freezes-out onto
the solids to account for the truncation. This water vapor depletion mechanism
by turbulent diffusion is related to the ``cold-finger effect'' of
\citet{StevensonLunine88}, but working in the vertical direction. Using
Monte-Carlo simulations of dust/vapor dynamics, \citet{RosJohansen13}
and \citet{Krijt16} indeed found rapid depletion of water vapor in the surface layer of 
inner PPDs near the water ice line. They further showed that the
depletion process strongly promote grain growth, and hence
planetesimal formation.

In this paper, we apply the picture of turbulent-diffusion mediated
volatile depletion to CO. Note that this picture has also been invoked by
\citet{Kama2016} recently to interpret carbon depletion based on ALMA
observations of CO and [CI] lines from two PPDs.
We focus on the outer regions of PPDs ($\gtrsim10-30$AU), which are
where most of the CO mass resides, and are also where significant CO
depletion has been observationally measured. Compared to the inner disk, the outer disk is 
characterized by much lower gas density and much longer dynamical
timescales. Correspondingly, dust
grains of the same size are more loosely coupled to the gas in the outer disk, and settle
more strongly towards the midplane, which, as this paper shows, has a large impact on volatile depletion.

In this work, we present a simple semi-analytical model for the evolution
of CO abundances to quantify the efficiency of gas-phase CO depletion. It
incorporates dust settling, turbulent diffusion, adsorption (freeze-out)
and thermal desorption (sublimation) processes. We do not attempt to model
the entire disk in full scale, but restrict
ourselves to a simple one-dimensional (1D) model in the vertical dimension.
Our goal is to demonstrate and clarify the relevant physics, which can be
incorporated into more sophisticated models in the future.
We highlight that the level of disk turbulence we adopt is motivated from
recent gas dynamic simulations of the outer PPDs \citep{Simon2013,Bai2015}  that take into account more realistic disk physics:
the level of turbulence in the outer disk is layered, with strong turbulence
in the warm disk surface layer and much weaker turbulence in the midplane
region (see Section \ref{sec:model} for more details).
We will show that this layered structure of turbulence facilitates the
depletion of gas-phase CO.

We describe the basic picture of turbulent-diffusion mediated CO depletion, as well as
our physical model in Section \ref{sec:model}. Model results presented in Section
\ref{sec:result}. In Section \ref{sec:con}, we summarize, and discuss the caveats and
applications.

\section{Model Description}\label{sec:model}

We are mainly concerned with the outer regions of PPDs, which are characterized
by low temperatures and weak turbulence in the midplane, and higher temperatures
and levels of turbulence in the disk surface layers. In Figure \ref{fig:PPDs},
we illustrate the basic picture of turbulent-diffusion mediated CO depletion,
and elaborate below.

Due to external heating by stellar UV and X-rays, PPDs exhibit a vertical
temperature gradient. Disk temperature is sufficiently low in the midplane
($\lesssim20$K, e.g., \citealp{Bisschop_etal06,Qi_etal13}) for CO to freeze out onto dust
grains, while disk surface is much warmer so that CO remains largely in gas
phase (e.g., \citealp{Glassgold04,KampDullemond04,Walsh_etal10,Rosenfeld_etal13}). This
dividing line for freeze-out is sometimes referred to as the ``atmospheric snow
line'', which is approximately two horizontal surfaces above/below the midplane.
The radial snow line, which is more commonly known, corresponds to the location
in the midplane region where temperature transitions through the CO freeze-out
temperature. It smoothly joins the atmospheric snow line towards larger radii,
as shown in Figure \ref{fig:PPDs}. We consider the regions beyond the
radial CO snow line, where only atmospheric snow lines are present.

Another key ingredient in the picture is the level of turbulence in PPDs.
Due to the weakly ionized nature of PPD gas, its dynamics suffers from strong
non-ideal magnetohydrodyamic effects (see \citealp{Turner_etal14} for a review),
which can significantly weaken, or suppress turbulence generated from the
magnetorotational instability (MRI, \citealp{BalbusHawley91}).
The gas dynamics in the outer PPD is relatively simple because ambipolar
diffusion (AD) is the only dominant non-ideal MHD effect. It has been
shown that the MRI in the disk midplane is substantially damped
by AD, leading to weak turbulence \cite{BaiStone11}. On the other hand,
the disk surface layer is likely much better ionized due to
far-UV radiaition \citep{Perez-BeckerChiang11}, leading to strong MRI
turbulence \citep{Simon2013,Bai2015}.

The basic mechanism of turbulent-diffusion mediated CO depletion is easily
explained: freeze-out of CO in the cold midplane creates a sink in gas-phase
CO, leading to a diffusive flux of CO gas into the midplane due to turbulent
mixing, which then quickly freezes out.

One caveat in the above picture is that the role of turbulence is two-fold.
Besides driving a diffusive flux of gas-phase CO down to the midplane,
turbulence can also potentially bring the dust to the warmer disk surface.
This would allow mantle-phase CO to sublimate and be released back to the gas
phase. If this process is efficient, it may largely cancel the aforementioned
CO depletion process. 
In a scenario where the disk midplane region is only very weakly turbulent,
it is possible that vast majority of the dust grains are strongly settled
towards the midplane and never rise to the surface, then the gas-phase CO
would be systematically depleted.

\subsection{Disk model and Dust Properties}\label{sec:dust}

\begin{figure}[!ht]
\centering
\includegraphics[width =0.47\textwidth]{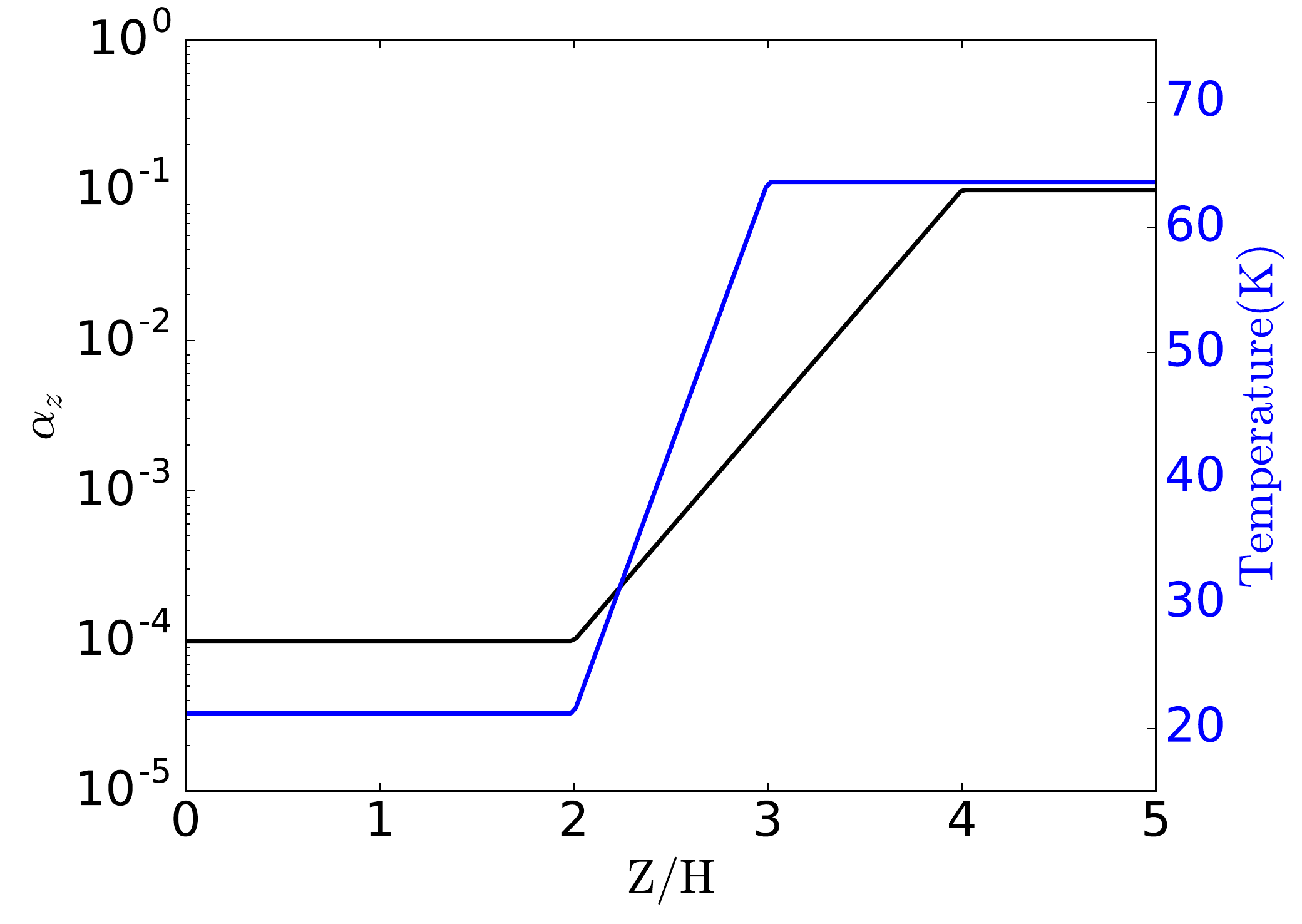}
\quad
\includegraphics[width =0.45\textwidth]{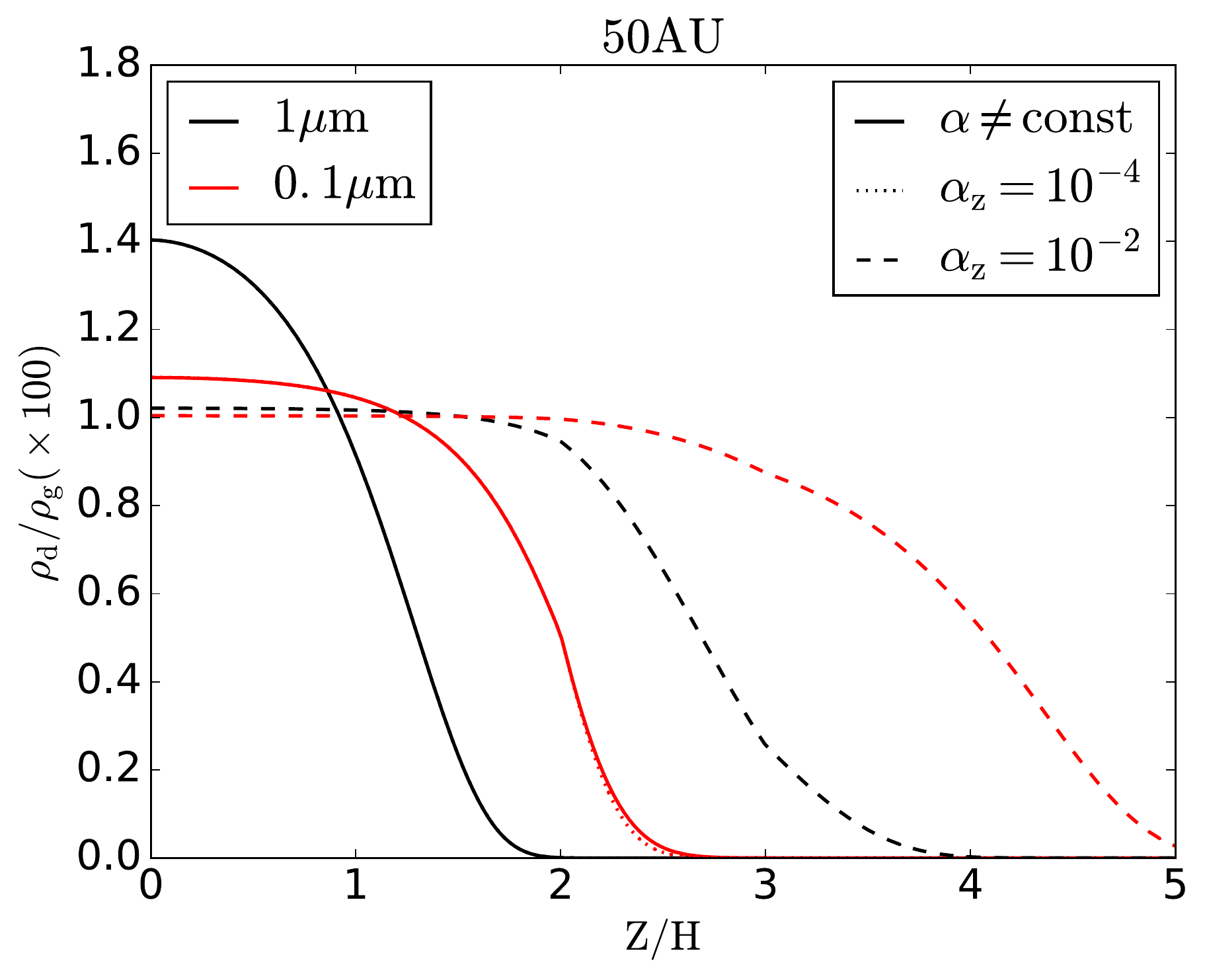}
\caption{Top: Vertical profile of the turbulent diffusion coefficient
$\alpha_z$ adopted in this work and vertical temperature profile at $\rm 50AU$.
Bottom: vertical profiles of single-sized 1$\mu$m (black) and $0.1\mu$m (red) dust at
$\rm 50AU$ with layered turbulent diffusivity profile (solid lines, $\alpha_z$ from
the top panel), constant $\alpha_z=10^{-4}$ (dotted lines) and constant
$\alpha_z=10^{-2}$ (dashed lines). Dust to gas mass ratio is $10^{-2}$. }
\label{fig:alpha}
\end{figure}

Motivated by observations, we consider a thin disk model with surface density
$\Sigma (r) = 500r_{\rm{AU}}^{-1}\ \mathrm{g\ cm^{-2}}$ and a midplane temperature profile $T_{\mathrm{mid}}(r)= 150r_{\rm{AU}}^{-1/2} \mathrm{K}$ 
where $r_{\rm{AU}}$ is disk radius measured in AU \citep{AndrewsWilliams07,Andrews2009}.
We focus on the outer regions of PPDs at radii $R\gtrsim30$ AU and choose a fiducial
radius of $R=50$AU, which is well outside of the radial CO snowline.
Disk vertical temperature profiles are poorly constrained observationally
(e.g. \citet{Rosenfeld_etal13}), and for simplicity we assume that the temperature is
$T=T_{\rm mid}(r)$ within two disk scale heights ($2H$, with $H$ measured at disk
midplane)
\footnote{Disk scale height $H$ is defined as $H = c_s/\Omega_K$, where $\Omega_K$
is the midplane Keplerian frequency, $c_s = \sqrt{k_B T_{\rm mid}(r)/\mu m_p}$ is
the midplane sound speed, with mean molecular weight taken to be $\mu=2.34$.}.
It increases linearly to $3 T_{\rm mid}(r)$ in one scale height and remains at
$3T_{\rm mid}(r)$ beyond $z=\pm3H$ (see Figure \ref{fig:alpha}a). Vertical density
structure is then simply determined from hydrostatic equilibrium.
To mimic the results of the layered turbulence profile, we employ a toy $\alpha_z$
profile where it increases exponentially from $10^{-4}$ starting at $2H$ to
$10^{-1}$ at $4H$ (see Figure \ref{fig:alpha}a), which is motivated from the vertical
turbulent velocity profiles from the outer PPD simulations of \citet{Bai2015}.

Dust grains interact with the gas via aerodynamic drag, with drag force given by $F_{\rm drag} =m_d\abs{\Delta\mathbf{v}}/t_{\rm stop}$,
where $m_d$ is dust mass,  $\Delta\mathbf{v}$ is the relative velocity between
dust and gas. In the outer disk, gas drag is in the Epstein's regime
\citep{Epstein24}, with stopping time $t_{\rm stop}$ given by
$t_{\rm stop}=\rho_sa/\rho_gc_s$, where $\rho_g$ is gas density, $\rho_s = 2\rm g/cm^2$ is the
dust solid density, $a$ is dust grain size.
It is more convenient to use the dimensionless stopping time
$\tau_s=t_{s}\Omega_{K}=\rho_s\Omega_K a/\rho_gc_s$. 

In the vertical dimension, dust mass density $\rho_d$ satisfies the continuity equation
\citep{TakeuchiLin2002}
\begin{equation}\label{eq:continuity}
\frac{\partial \rho_d}{\partial t} + \nabla\cdot(\rho_dv_d+j_d) = 0\ ,
\end{equation}
where $v_d$ is the dust settling velocity, $j_d$ is dust diffusive flux
given by
\begin{equation}\label{eq:diffdust}
j_d = -\frac{\rho_g D_z}{S_c}\frac{\partial}{\partial z}
\left( \frac{\rho_d}{\rho_g}\right)\ .
\end{equation}
In the above, $D_z$ is the diffusion coefficient, which we parameterize as
$D_z=\alpha_z c_s H$ \citep{ShakuraSunyaev1973}. 
 The Schmidt number $S_c$
represents the strength of coupling between dust and the gas, which is
approximated as $1+\tau_s^2$ for turbulent eddy time $\sim\Omega_K^{-1}$
\citep{Youdin2007}. The settling velocity is given by
\begin{equation}
v_d=-\Omega_Kz\tau_s/S_c\ .
\end{equation}
The extra factor $S_c$ guarantees that in the limit $\tau_s\ll 1$, $v_d$ is
simply the dust terminal velocity. In the opposite limit $\tau_s>1$, dust
particles undergo damped oscillation about the midplane with mean settling
velocity $\sim 1/\tau_s$ that is also captured by the expression.

In steady state, the solution to Equation
(\ref{eq:continuity}) can be expressed as
\begin{equation}\label{eqn:dustv}
\left(\frac{\rho_d}{\rho_g}\right)_{z=z_0} = \left(\frac{\rho_d}{\rho_g}\right)_{z=0} 
\mathrm{exp}\left[\int_0^{z_0}-\frac{\tau_s(z^\prime)z^\prime}{\alpha_z(z^\prime) H^2}dz^\prime\right]\ .
\end{equation}
Note that both $\tau_s$ and $\alpha_z$ depend on $z$, and substantial dust
settling is realized around the height where $\tau_s(z)\sim\alpha_z(z)$.
For constant $\tau_s$ and $\alpha_z$, the dust distribution becomes a
Gaussian with scale height is $\sqrt{\alpha_z/\tau_s}H$, recovering the
result of \citet{Youdin2007} {\bf in the limit of $\alpha_z<\tau_s$}.
 
In Figure \ref{fig:alpha}b, we show the dust density profile at $50\rm AU$,
and compare with constant $\alpha_z$ profiles. For constant $\alpha_z=10^{-4}$,
even sub-micron sized dust settle to within $\pm2H$.
For stronger turbulence $\alpha_z=10^{-2}$, however, dust particles are
stirred to well above $\pm3H$ from the midplane. Using our adopted
layered $\alpha_z$ profile, the results are almost identical to the constant
$\alpha_z=10^{-4}$ case, because the layered profile shares the same
$\alpha_z$ in the bulk of the dust layer.

The above calculation assumes a single dust size, but in reality, dust grains
collide with each other, resulting in a size distribution from coagulation and
fragmentation \citep{Birnstiel2011}. In this work, We first consider a single
grain size to clarify the most important physics. We then proceed with a more
realistic size distribution in the form of $n_d(a)\propto a^{\beta}$ where
$n_d(a)da$ is the grain number density between size $a$ and $a+da$. We take
$\beta=-3.5$ as in the standard MRN size distribution \citep{MRN1977}. The largest grain size is fixed to be $1$ cm, and the smallest grain size $a_{\rm min}$ is chosen
to be $0.1\mu$m, $1\mu$m or $10\mu$m. The $0.1\mu$m size is typical from dust coagulation
models (e.g., \citealp{Birnstiel2011}), and larger sizes are chosen to mimic the effect of grain growth.\footnote{Although the dust coagulation model of \citet{Birnstiel2011} gives a set of broken
power-laws instead of a single power-law, we will show in
\S \ref{sec:dustdistribution} that the details of the size distribution
do not affect our main results.} The total dust mass ratio is set to $10^{-2}$, and calculations are conducted for individual dust size bins.

In our calculations of volatile evolution, the dust density profile is fixed
based on the steady state profile (\ref{eqn:dustv}). Although the steady-state
assumption may not perfectly hold, we note that the only essential requirement
of the mechanism we study is that the bulk of the dust settles to the cold midplane
layer within the atmospheric snow line (in our model within $\pm2H$ about the midplane).
Since the stopping time of grains increases towards disk surface, the timescale
for dust to settle to within 2H is much shorter. Although they do not
necessarily achieve a fully steady state distribution, we find that for grain
size $\gtrsim 0.1\mu$m, the settling timescale above $z\sim 2H$ is within 1 Myr,
which suffices for our purpose.

\subsection{Volatile Adsorption and Desorption on Grains}\label{sec:chem}

Volatiles interact with dust grains \textit{via} adsorption/desorption.
The adsorption rate of gas phase CO onto grains of size $a$ is given by
\begin{equation}\label{eq:ad}
    R_{ad}(a)\approx v_{th}\pi a^2 n_d(a)n_{\mathrm{co}[g]},
\end{equation}
where $v_{th}$ is the thermal velocity of CO, $n_{\mathrm{co}[g]}$  is the
gas phase CO number density.
The sticking coefficient is approximately 1 and is left out
\citep{Bisschop_etal06}. Total adsorption rate is a summation over the
grain size distribution. 

For this work, it suffices to only consider thermal desorption, whose
rate is given by \citep{Hasegawa1992} 
\begin{equation}\label{eq:desorp}
R_{des}(a)= n_{\mathrm{co}[m]}(a)\nu_0 \mathrm{exp}[-E_D/k_B T]/\mathrm{Max}(1,N_{\rm layer})
\end{equation}
where $n_{\mathrm{co}[m]}(a)da$ is number density of adsorbed CO molecules (mantles)
on the surface of dust grains between size $a$ and $a+da$. 
$\nu_0 = (2 n_s E_D/\pi^2 m_{\mathrm{co}})^{1/2}$ is the characteristic vibration 
frequency with surface density of adsorption sites 
$n_s\simeq1.5\times10^{15}$cm$^{-2}$, $m_{\mathrm{co}}$ is CO mass and we take
the CO binding energy $E_D=1150$K. While the exact value of binding energy depends
on grain surface properties and can be as low as $\sim850$K \citep{Oberg_etal05},
it does not affect the physics we present here. The extra factor $N_{\rm layer}= n_{\rm co[m]}(a)/4\pi a^2n_sn_d(a)$ is the ice layer on dust grains. Since for ices thicker than monolayer, only the top molecular layer is available for desorption.
Note that we assume that volatile freeze-out does not appreciably change the dust
grain sizes.

\subsection{1D Evolution model}\label{sec:eqn}

We solve time-dependent 1D evolution equations for gas-phase CO and the
mantle phase CO for individual dust size bins. 
The equation for the gas phase CO number density reads
\begin{equation}\label{eqn:evog}
 \frac{\partial n_{\mathrm{co}[g]}}{\partial  t} =- \nabla\cdot
 j_{\mathrm{co}[g]} + \sum_a\left[ R_{des}(a)- R_{ad}(a)\right]da\ ,
\end{equation} 
where the CO diffusive flux $j_{\mathrm{co}[g]}$ is given by
\begin{equation}
  j_{\mathrm{co}[g]} = -n_g D_z
  \frac{\partial}{\partial z}\left( \frac{n_{\mathrm{co}[g]}}{n_g}\right)\ ,
\end{equation}
with $n_g$ being gas number density.

The evolution for mantle phase CO associated with grain size $a$ is given by
\begin{equation}\label{eqn:evom}
\begin{split}
\frac{\partial n_{\mathrm{co}[m]}(a)}{\partial t}
 =&-\nabla\cdot\left[n_{\mathrm{co}[m]}(a)v_d+j_{\mathrm{co}[m]}(a)\right]\\
  &+R_{ad}(a) -R_{des}(a)\ ,
\end{split}
\end{equation}
where the diffusive flux follows from dust dynamics
\begin{equation}\label{eq:diffman}
j_{\mathrm{co}[m]}(a)= -\frac{n_g D_z}{S_c}\frac{\partial}{\partial z}\left(\frac{n_{\mathrm{co}[m]}(a)}{n_g}\right)\ .
\end{equation}
 In this formulation, we assume the mantle-CO abundance at each height for
each grain size is single-valued, representing the mean value. In reality,
mantle-phase CO abundance on each grain can be different depending on individual
grain trajectories. However, such further complication can be essentially
absorbed into the desorption prescription via Equation (\ref{eq:desorp}),
and should not affect the physics we address in this paper.

At $t=0$, we assume CO molecules are all in the gas phase and are well mixed with the bulk gas. Prescribing fixed turbulence and dust background profile as discussed earlier,
the equations are evolved for 1 Myrs.

\begin{figure*}[!ht]
\centering
\includegraphics[width =0.9\textwidth]{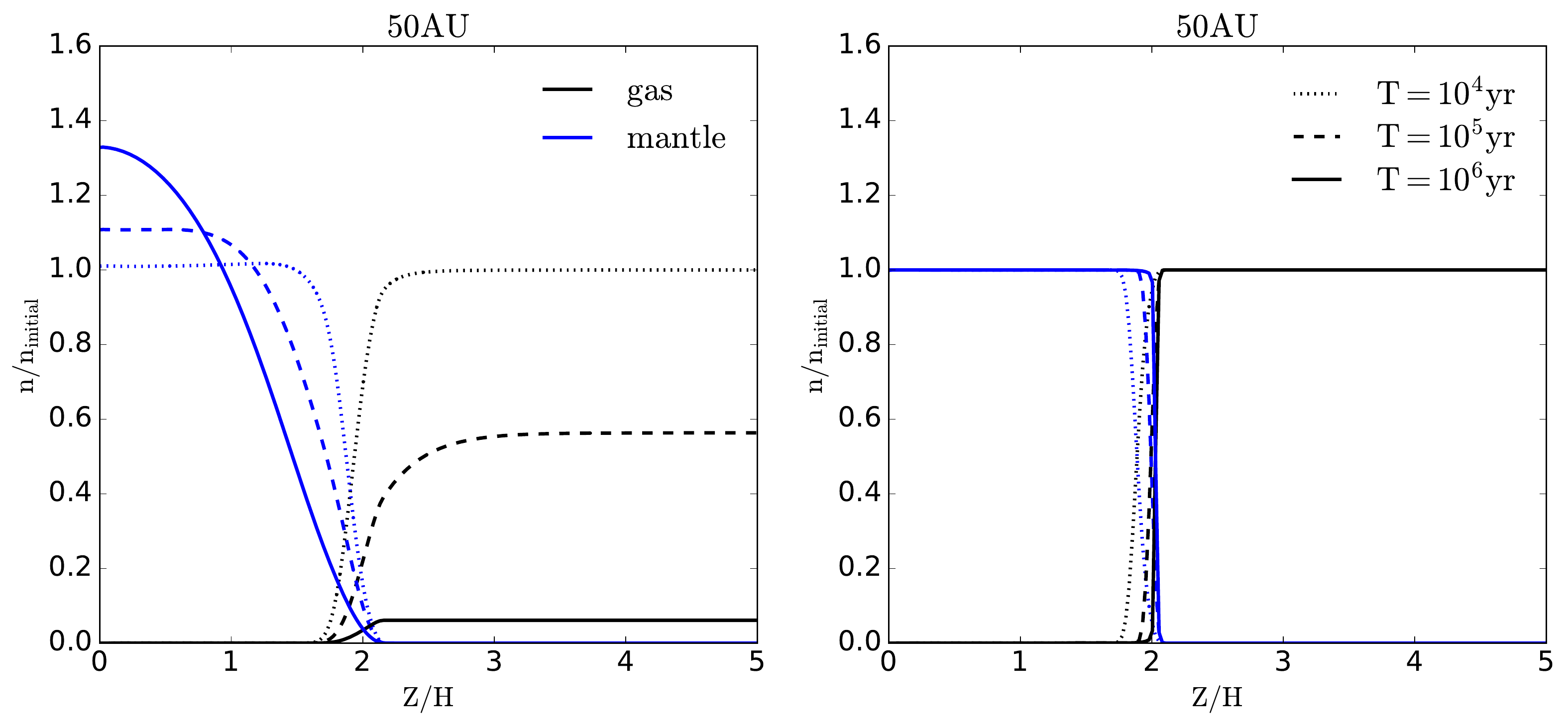}
\caption{Time evolution of gas phase (black lines) and mantle phase (blue lines) CO number densities
(normalized by initial gas phase CO number density) assuming single sized
$1\mu m$ dust in the layered $\alpha_z$ profile at $\rm 50AU$. Left/right panel corresponds to calculations that include/exclude turbulent
diffusion of CO. Different linestyles correspond to $10^4$ yrs (dotted lines), $10^5$ yrs (dashed lines), $10^6$ yrs (solid lines), respectively.}
\label{fig:evo}
\end{figure*}

\section{Results}\label{sec:result}

We discuss the results of our calculations in this section.
We start from the simplest example with a single grain size,
and then consider the more realistic case with a grain size
distribution.

\subsection{Time evolution}

In Figure \ref{fig:evo}, we show the time evolution of gas/mantle-phase
CO number densities for fixed grain size $a=1\mu m$ at $\rm 50AU$ using
layered $\alpha_z$ profile. At the start of the model, CO freezes out
rapidly in the midplane, while it remains almost completely in gas phase
in the warmer surface layer, with the atmospheric snow line located at
$z\sim2H$. This sharpness of the transition is the result of extremely
sensitive temperature dependence of the freeze-out process.

Most existing chemistry models ignore turbulent diffusion,
which are equivalent to removing the divergence terms in
Equations (\ref{eqn:evog}) and (\ref{eqn:evom}).
In this case, the results correspond to the right panel of Figure
\ref{fig:evo}, where the gas phase CO abundance is entirely determined by
the adsorption/desorption process of the initial CO reservoir, and shows
very little time evolution.

Including turbulent diffusion (left panel of Figure \ref{fig:evo}), we
see that CO at the surface layer is gradually depleted over the timescale
of $10^{5-6}$ years, accompanied by the enhancement of mantle
phase CO in the midplane. This confirms the picture
of we outlined at the beginning of Section \ref{sec:model}. By 1Myr,
we see that the gas phase CO in the surface layer is depleted by a factor
of more than $10$, which is comparable to the observed level of CO
depletion in PPDs.

\begin{figure*}[!ht]
\centering
\includegraphics[width =0.43\textwidth]{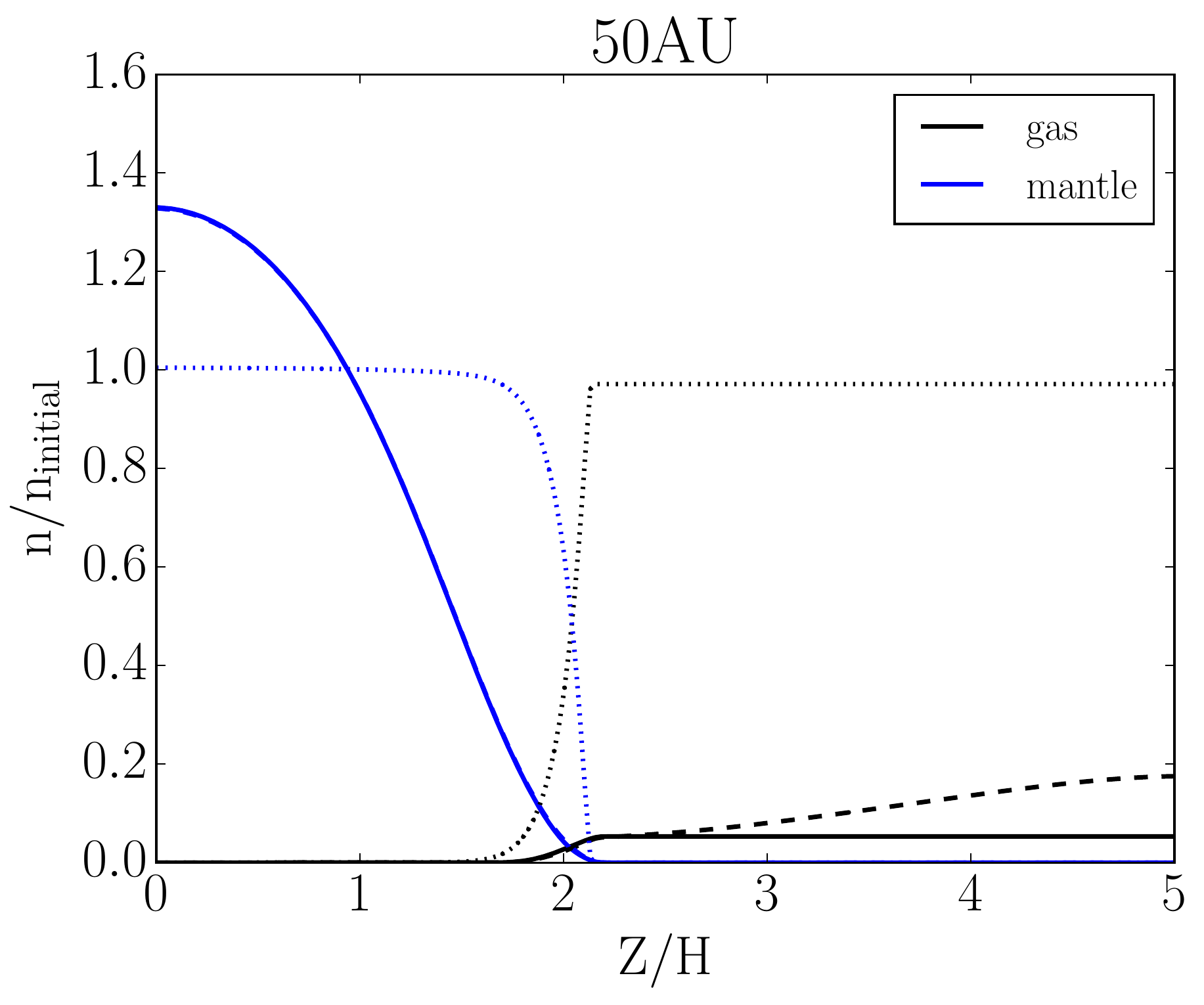}
\quad
\includegraphics[width =0.43\textwidth]{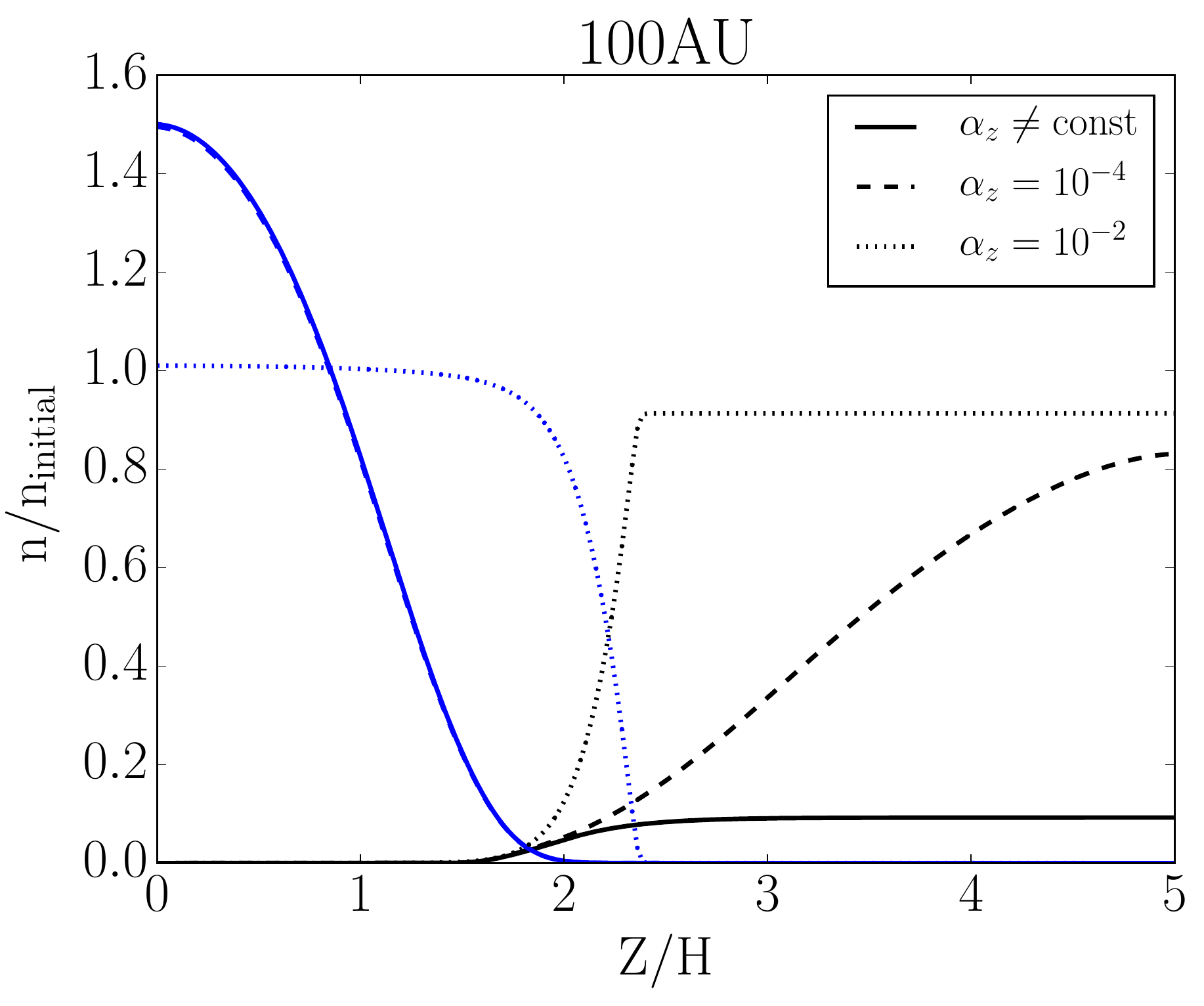}
\quad
\includegraphics[width =0.43\textwidth]{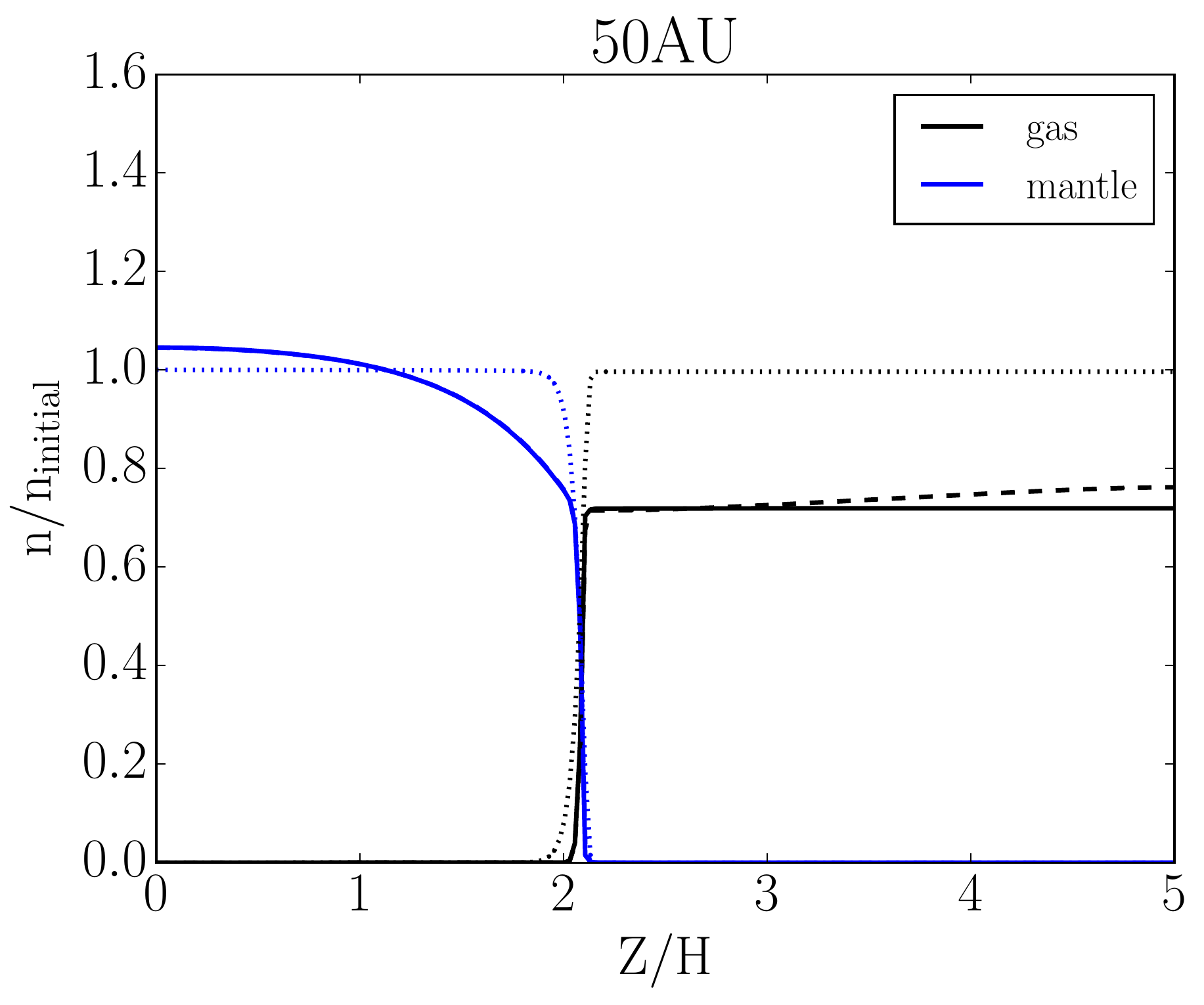}
\quad
\includegraphics[width =0.43\textwidth]{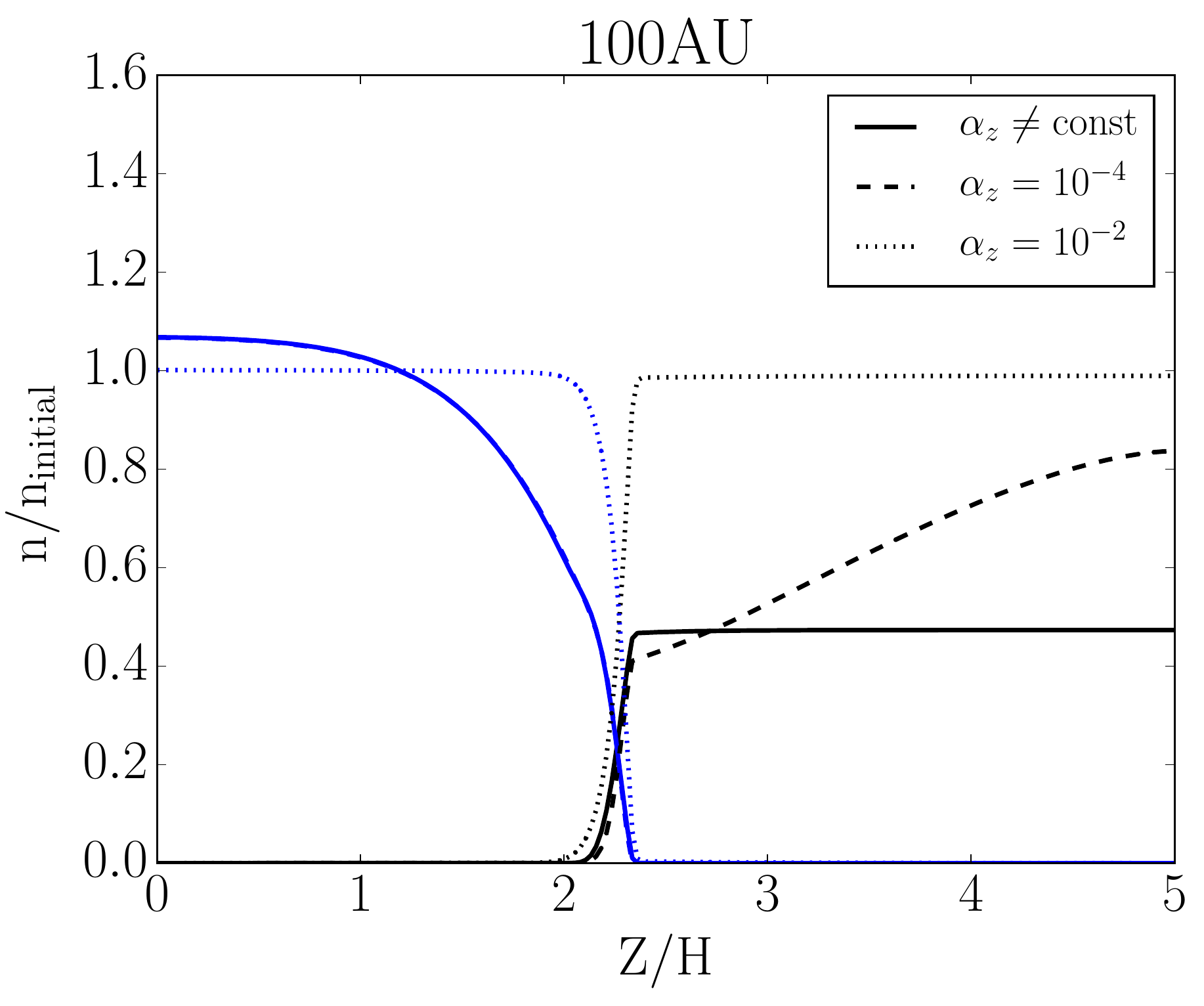}
\caption{Vertical profiles of gas phase (black) and mantle phase (blue) CO number densities after $\rm 1Myr$ evolution (normalized to initial 
CO number density) at $\rm 50AU$(left) and $\rm 100AU$(right) with layered 
$\alpha_z$ profile (solid), constant $\alpha_z=10^{-4}$ (dashed line) and constant $\alpha_z=10^{-2}$ (dotted line). Grain size is fixed at $0.1\mu m$ on the top panels and $1\mu m$ on the bottom panels.}
\label{fig:comp}
\end{figure*}

Finally, we note that in this calculation, the gas and mantle phase CO
profiles have reached an approximate steady-state right around 1 Myrs, where
the downward flux of gas-phase CO balances the upward flux of mantle-phase CO.
The level of gas-phase CO depletion in steady state is mainly determined by
the dust density gradient across the atmospheric snow line, reflecting the
level of dust settling. Typically, this steady state is achieved over a few
times the turbulent diffusion timescale at the atmospheric snow line.

\subsection{Conditions for Strong CO Depletion}\label{sec:cond}

The example in the previous subsection is to a certain extent optimized.
We now discuss under what conditions CO depletion is efficient
by conducting calculations with different grain sizes, disk radius
and turbulence profiles. The results after evolving for 1 Myr are shown
in Figure \ref{fig:comp}.

We find that the most stringent requirement to achieve strong CO depletion
is a sufficiently weak turbulence in the midplane so that vast majority of
dust grains can settle within the atmospheric snow line.
This condition is demonstrated from two aspects in Figure \ref{fig:comp}. First, all calculations with strong turbulence (constant $\alpha_z=10^{-2}$)
show essentially no CO depletion. This is because strong turbulence stirs up
dust grains to the warmer disk surface, where CO mantles desorb and return
to the gas phase. In steady state, there is an efficient CO circulation with
downward turbulent mixing in the gas phase and upward turbulent diffusion of
icy grains. The cancellation of the two effects lead to an outcome
that is similar to the case without turbulent mixing. On the other hand,
with weak midplane turbulence ($\alpha_z=10^{-4}$ or the layered profile case),
all calculations show certain level of CO depletion in the disk surface.
Second, more loosely coupled grains lead to more significant depletion, and
vice versa. Our calculation with $a=0.1\mu$m show only low level of CO
delpetion, because these grains only partially settle to the cold
midplane region, and instead regularly return to the warmer disk surface.

Too much settling slows down CO depletion, however.
If dust grains concentrate too close toward disk midplane,
the CO depletion rate is reduced because of decreased surface area and
therefore longer freeze-out timescales at the CO atmospheric snowline.
Upon reaching steady state, 
depletion of gas phase CO would
be more complete, but this may never be achieved dependent on the disk life time. 

The competing effects of settling on CO depletion after 1Myr is readily seen when comparing models run at 50 and 100 AU. Compared to 50 AU, grains at 100 AU of a fixed
size are 
more loosely coupled (having larger $\tau_s$). We see that
for $0.1\mu m$ grains, more CO depletion is achieved at 100 AU compared
to 50 AU, which is because they settle more. For $1\mu m$ grains,
the depletion is slightly less at 100 AU after 1 Myr because the larger
grains largely settle far below the atmospheric snowline, leading to a
longer CO depletion timescale, as discussed above. If we
continue to evolve the system for $\sim1-2$ Myrs, more CO depletion can
be achieved than in the 50 AU case.

Strong turbulence at the disk surface enhances gas-phase CO depletion.
The turbulent mixing timescale is given by
$t_{\rm mix} \simeq H^2/D_z\sim(\alpha_z\Omega_K)^{-1}$.
Freeze-out of CO brought from disk surface primarily occurs around the
atmospheric snow line, in our case at slightly above $z=2H$. For constant
$\alpha_z=10^{-4}$, the mixing timescale is the same at all heights. As a
result, depletion of CO occurs first at the atmospheric snow line, and
then propagates towards disk upper layer, leading to a
slow depletion towards disk surface. 
With a layered $\alpha_z$ profile, much more efficient turbulent mixing
at the disk surface homogenize the gas-phase CO distribution there. As a
result, CO depletion proceeds almost simultaneously through the entire
disk surface column.

\begin{figure*}[!ht]
\centering
\includegraphics[width =0.45\textwidth]{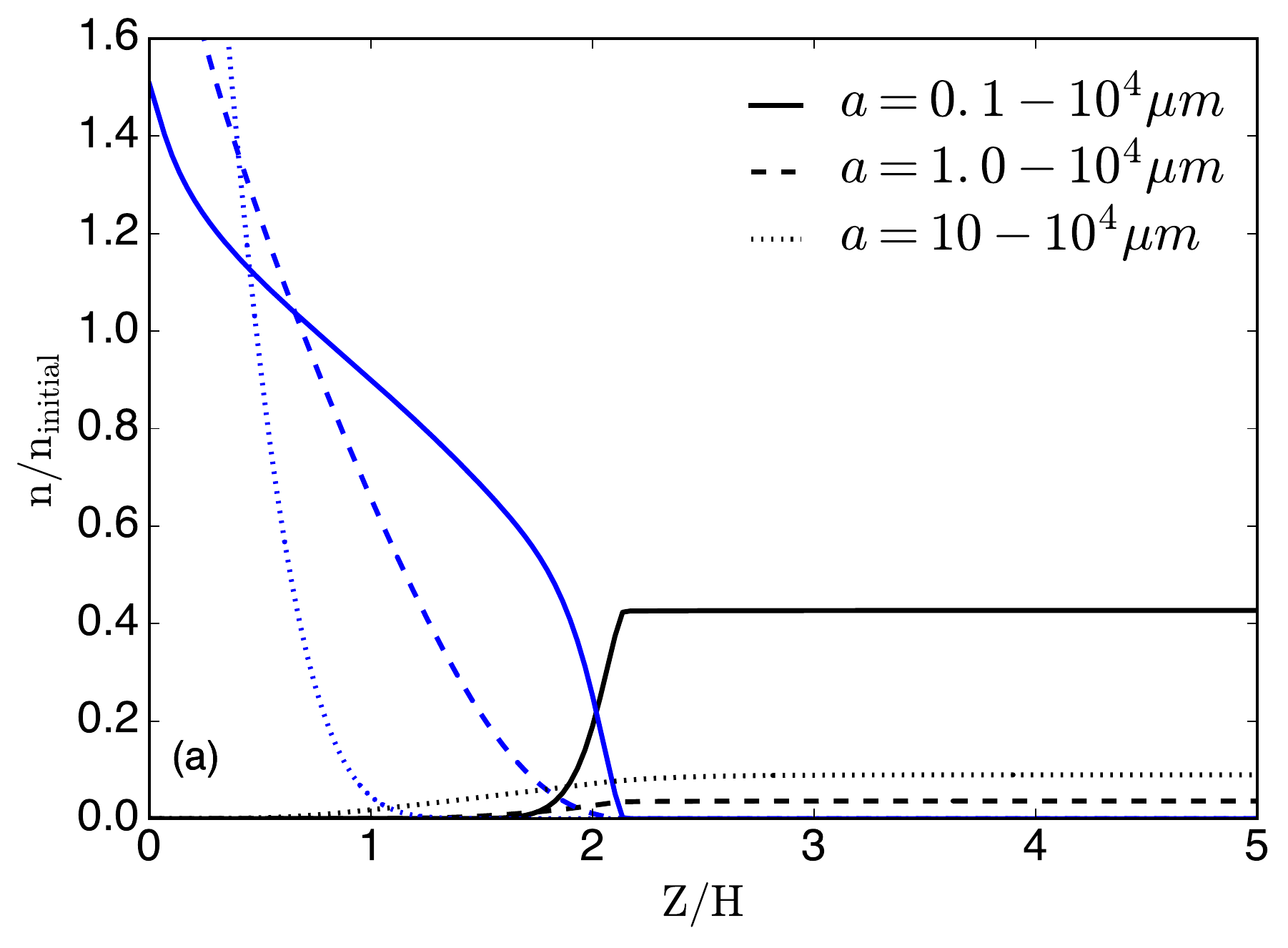}
\quad
\includegraphics[width =0.45\textwidth]{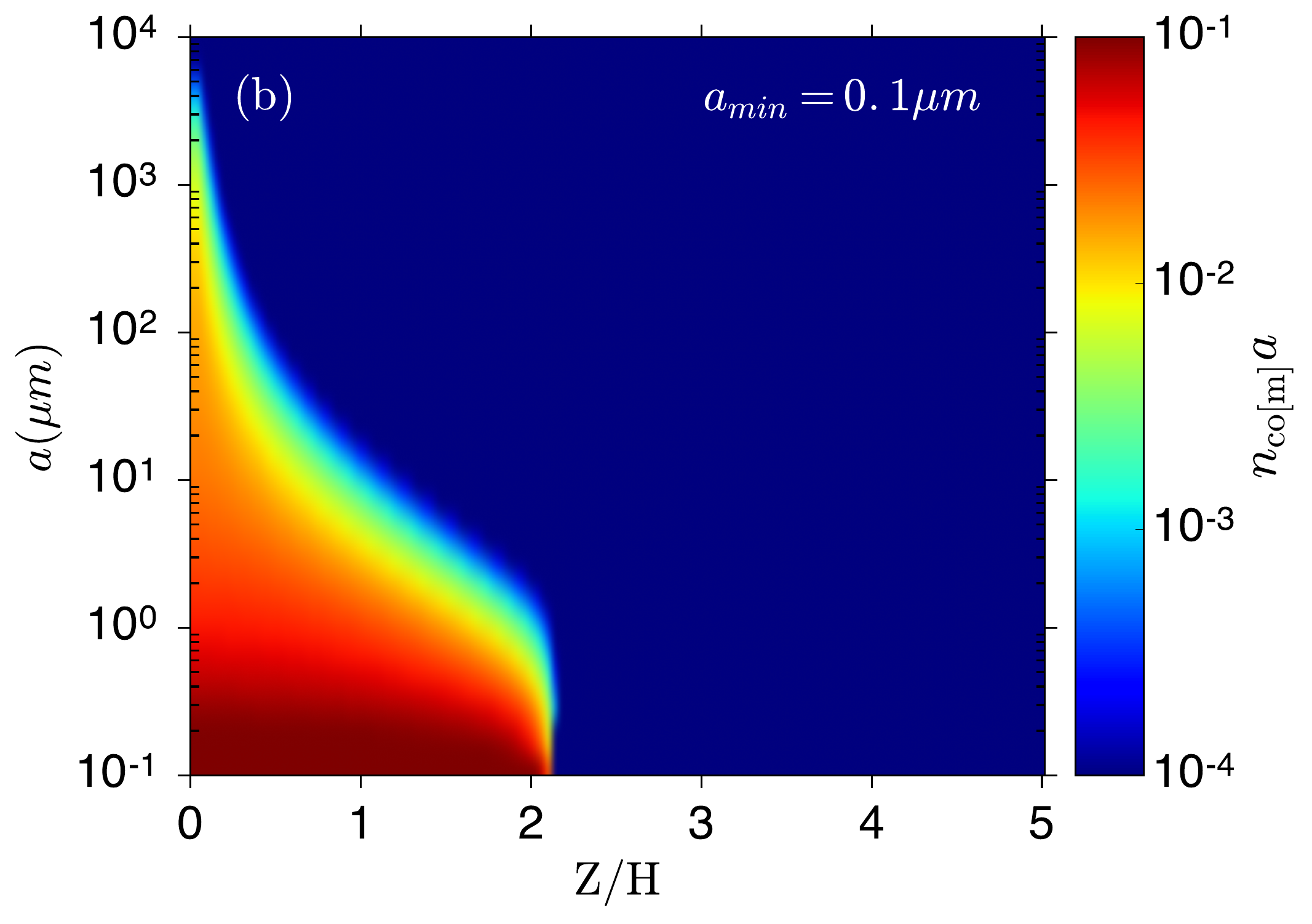}
\includegraphics[width =0.45\textwidth]{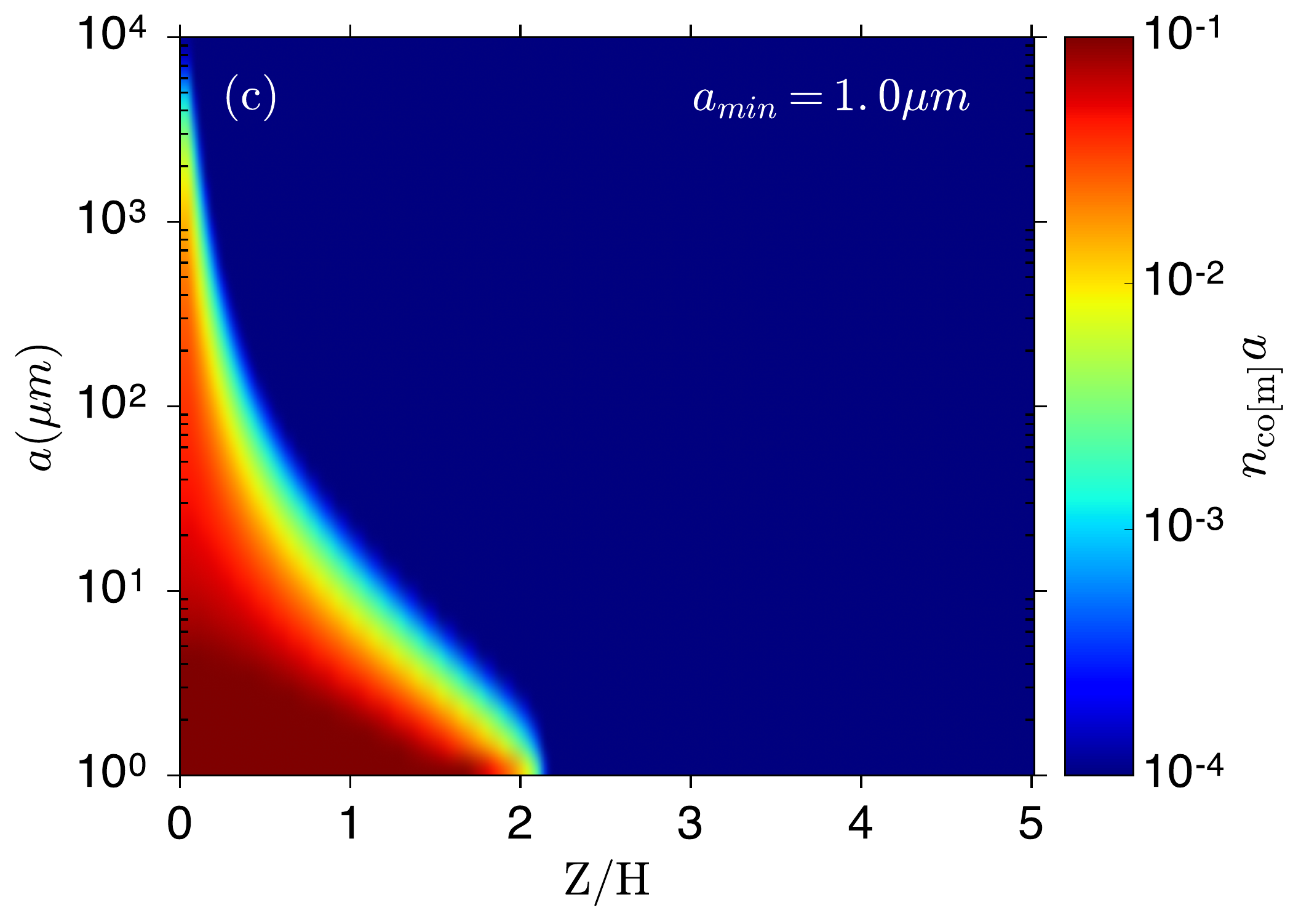}
\quad
\includegraphics[width =0.45\textwidth]{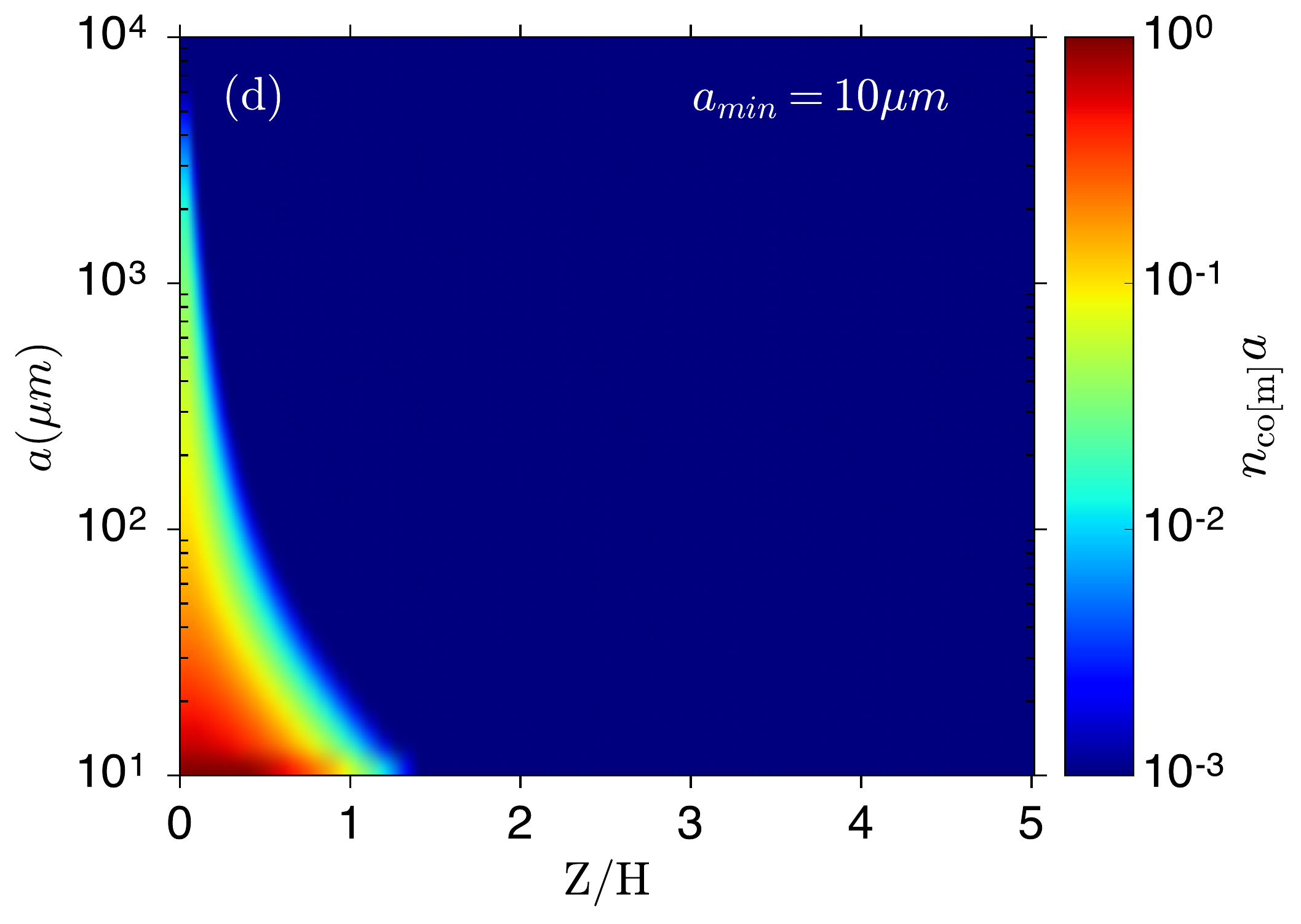}
\caption{(a): Gas phase (black) and total mantle phase (blue) CO number densities
after 1 Myr evolution (normalized to initial CO number density) at $\rm 50AU$ with layered $\alpha_z$ profile and a dust size distribution in
the size range of $a =0.1\mu m-10^4\mu m$ (solid), $a=1\mu m -10^4\mu m$ (dashed) and $a = 10\mu m -10^4\mu m$ (dotted),
respectively. (b)-(d): mantle phase CO number density as a function of dust size and altitude for the three calculations considered in the (a). Color represents is the CO mantle number density per logarithmic dust size scale $an_{co[m]}(a)$ (has the dimension as number density), normalized to initial CO number density, after 1 Myrs of evolution. (b)-(d) correspond to calculations with smallest grain size of $a_{\rm min}=0.1\mu$m, $1\mu$m and $10\mu$m, respectively.}

\label{fig:multi}
\end{figure*}

\subsection{Importance of dust distribution characteristics}\label{sec:dustdistribution}

In this subsection, we consider the effects of dust size distributions with the three size
ranges mentioned in Section \ref{sec:dust}. They differ in the smallest
grain size $a_{\rm min}$. 
Theoretically, we expect the smallest grains to play the dominant role in
the outcome of CO depletion. This is because the smallest grains generally
dominate the total dust surface area, and are more suspended and therefore 
directly accessible to the fresh CO molecules brought down from the disk
surface. Larger grains contribute less or even negligibly to CO freeze-out
because they have much smaller total surface area. 
Also, they often settle well below the atmospheric snow line  
and therefore have little direct access to the gas phase CO.

Figure \ref{fig:multi}(a) shows gas phase and total mantle
phase CO number densities after 1 Myr evolution with the three different size
distributions. We can see in cases with $a_{\rm min}=0.1\mu$m
(solid line) and $1\mu$m (dashed line), the overall level of CO depletion 
is comparable with their single grain size ($0.1\mu$m and $1\mu$m)
counterparts shown in Figure \ref{fig:comp}, which confirm our expectations.
Increasing $a_{\rm min}$ further to $10\mu$m  leads to
slower depletion (dotted line). 
In this case, even the smallest
grains are large enough to settle well below the atmospheric snow line (see Section \ref{sec:cond}),
leading to longer depletion timescale.

Figure \ref{fig:multi}(b)-(d) show the distribution of
mantle phase CO as a function of grain size and vertical height.
We see that most of the mantle phase CO are in the smallest grains, which
confirms our previous discussion that CO depletion is dominated by the
smallest grains. We note that $a\sim1\mu$m is a critical grain size in our
calculations. For $a\lesssim1\mu$m, 
a considerable fraction of grains stay outside the atmospheric snowline. 
In the case of $a_{\rm min}=10\mu$m, most mantle-phase CO
is well within the atmospheric snow line, which simply reflects the deficit
of grains at the atmospheric snow line. We see that the
$a_{\rm min}=1\mu$m case leads to fastest CO depletion. It suggests that
CO depletion is the most efficient when the population of grains containing
most of the surface area (smallest grains) settle just within the atmospheric
snow line.

 Note that we have assumed a fixed grain size in our calculations.
In reality, coagulation of small grains into larger grains,
followed by the settling of larger grains may increase the
mantle-phase CO abundances on larger grains than shown in
Figure \ref{fig:multi}.

Overall, with a grain size distribution, the rate of CO depletion is comparable
with the case of single-sized grain at smallest size. Over a 1 Myr timescale,
depletion of CO by a factor of $\sim2$ can be achieved for the more conservative
case with smallest grain size being $0.1\mu$m, and more complete depletion can
be achieved if grains grow bigger. We thus conclude that as long as the
conditions discussed in the previous subsection are satisfied,
turbulent diffusion is a robust mechanism for volatile depletion.

\section{Summary and Discussion}\label{sec:con}

In this paper, we have presented a simple semi-analytical model
which demonstrates that as a result of dust settling and
turbulent diffusion, CO in the warm surface layer of the outer
regions of PPDs are subject to turbulent diffusion into the cold
midplane and subsequent depletion.

The most important condition for turbulent-diffusion mediated CO
depletion is that midplane turbulence must be sufficiently weak so
that the bulk of the small grains that dominate
the surface area can settle within the atmospheric snow line. 
The process is
facilitated by stronger turbulence in the disk surface layer. Both
conditions are likely realizable in the outer regions of PPDs
\citep{PerezBeckerChiang11,Simon2013,Bai2015}.

Our results suggest that turbulent-diffusion 
likely contributes to the observed carbon (especially CO) depletion in PPDs,
particularly in the disk surface region (e.g., \citealp{Du_etal15}). Its
contribution depends on the level of midplane turbulence and grain size
distribution, and can be a factor of a few to more than one order of
magnitude. In reality, we expect that additional mechanisms also contribute
to carbon depletion. Conversion of carbon to complex organic molecules
likely yields a factor of a few of depletion over disk lifetime
\citep{Bergin2014,Yu2016}. Gas removal from disk wind likely contributes
another factor of two to a few to the reduced gas-to-dust mass ratio given
that wind mass loss rate is comparable to the mass accretion rate
\citep{Bai16}. Altogether, these processes are likely to be able
to account for a wide range (two orders of magnitude) of the CO depletion
factor, and/or the apparent gas-to-dust ratio inferred from observations.

While the mechanism of CO depletion in the outer disk discussed here is
similar to previous studies on the depletion of water vapor in the inner
disk \citep{Meijerink09,RosJohansen13,Krijt16}, there are important
differences. One may not directly apply our calculation results
to the inner disk, and vice versa.

We first note that the level of turbulence in
the inner region of PPDs is likely very weak yet highly uncertain. On the
one hand, the MRI is largely suppressed (e.g., \citealp{BaiStone13}),
yet some turbulence is needed to keep small grains suspended to match
the observed disk spectral energy distributions (SEDs), which may
be attributed to certain poorly understood hydrodynamic instabilities
(e.g., \citealp{Turner_etal14}). We choose to focus on the outer regions
of PPDs, where layered turbulence structure is likely a natural outcome
of the MRI operating in the surface FUV layer \citep{Perez-BeckerChiang11}.

Another important difference between the inner and outer regions of PPDs
is the timescale. The inner disk is characterized by fast collision and
grain growth timescales, and sub-micron grains are mainly the product of
destructive collisions of large grains. Recently, \citet{Krijt2016} showed
that because of their short collisional coagulation timescale, small grains
can be effectively trapped in the midplane region without diffusing to
disk upper layers. This effect can substantially enhance volatile depletion, and in the case of water, such depletion further enhances grain growth \citep{RosJohansen13,Krijt16}.

We have ignored the growth and collisional evolution of grains in our calculations. While this is less
self-consistent, we note that in the outer disk, grain growth is much slower
and is found to be drift-limited rather than fragmentation limited (e.g.,
\citealp{Birnstiel12}). Therefore, the smallest grain population in the
outer disk is mostly primordial rather than from collisional fragmentation. 
As we have mentioned in Section \ref{sec:dust}, it takes
$\sim1$Myr for $\sim0.1\mu$m grains to settle to within $\sim2H$ at
$\sim50$AU. We also estimate a similar $\sim$Myr time scale for collisions
between sub-micron grains at that height (which is more relevant for CO
depletion, instead of midplane). 
Therefore, because of the long timescale in the outer disk,
we expect grain growth to only play a minor role in the outer disk physics
discussed in this work.

As an initial effort, we focus on the physics of the mechanism using a
simple 1D model, which captures the essence of the problem. One important
limitation is that we have ignored the radial dimension, where grains
undergo radial drift, and the disk itself evolves over Myr timescales
\citep{TakeuchiLin2002}. We note that significantly improvement in our
knowledge about disk evolution is needed before more reliable dust transport
model can be made, especially given the prevalence of disk substructures
that has been realized in the recent years (e.g.,
\citealp{ALMA15,Nomura_etal16,Andrews_etal16,Zhang_etal16}). Regardless of the details
of radial dust transport, we expect our main conclusions to be robust as
long as the vertical profile of turbulence does not vary significantly with
radius in the outer disk.

Overall, our work has demonstrated the importance of incorporating
more realistic disk dynamics (i.e., turbulent diffusion) into models
of volatile evolution (e.g., \citealp{CieslaCuzzi06}).
The outcome would be important for determining, for instance, the
location of the volatile condensation fronts/snow lines
\citep{Oberg2011,Qi_etal13,Piso2016}, and volatile delivery to planets
which would affect the planets' bulk and atmospheric composition
\citep{Madhusudhan2011}. More generally, volatiles play an important
role in the overall disk chemistry \citep{HenningSemenov13}. As initially
pursued in \citet{SemenovWiebe11}, we expect future studies of PPD chemical
evolution to pay more attention to, and eventually benefit from
incorporating more realistic PPD gas dynamics.


We thank Chunhua Qi, Fred Ciesla, Til Birnstiel and Ted Bergin for
helpful discussions, and Mihkel Kama, Klaus Pontoppidan, Eugene
Chiang and an anonymous referee for useful comments that greatly improve the paper.
XNB acknowledges support from Institute for Theory and Computation,
Harvard-Smithsonian Center for Astrophysics. KI\"O acknowledges funding through a Packard Fellowship for Science and Engineering from the David and Lucile Packard Foundation.

\bibliographystyle{apj}
\bibliography{disk}
\label{lastpage}
\end{document}